\documentclass[twocolumn,aps,superscriptaddress,showpacs]{revtex4}
\usepackage[colorlinks=true,linktocpage=true,linkcolor=blue,citecolor=blue,allcolors=blue]{hyperref}
\usepackage{epsfig}
\usepackage{latexsym}
\usepackage{xspace}
\usepackage[utf8]{inputenc}
\usepackage{indentfirst}
\usepackage{enumerate}
\usepackage{color}
\usepackage{tabularx}

\usepackage{amsmath}
\usepackage{amssymb}
\usepackage[english]{babel}
\usepackage{url}
\begin{document}


\title{Application of artificial intelligence in the determination of impact parameter in heavy-ion collisions at intermediate energies}

\author{Fupeng Li}
\affiliation{College of Science, Zhejiang University of Technology, Hangzhou 310014, China}
\affiliation{School of Science, Huzhou University, Huzhou 313000, China}

\author{Yongjia Wang\footnote{Corresponding author: wangyongjia@zjhu.edu.cn}}
\affiliation{School of Science, Huzhou University, Huzhou 313000, China}

\author{Hongliang L\"u}
\affiliation{HiSilicon Research Department, Huawei Technoloies Co., Ltd., Shenzhen 518000, China}

\author{Pengcheng Li}
\affiliation{School of Nuclear Science and Technology, Lanzhou University, Lanzhou 730000, China}
\affiliation{School of Science, Huzhou University, Huzhou 313000, China}

\author{Qingfeng Li \footnote{Corresponding author: liqf@zjhu.edu.cn}}
\affiliation{School of Science, Huzhou University, Huzhou 313000, China}
\affiliation{Institute of Modern Physics, Chinese Academy of Sciences, Lanzhou 730000, China}

\author{Fanxin Liu}
\affiliation{College of Science, Zhejiang University of Technology, Hangzhou 310014, China}

\begin{abstract}
The impact parameter is one of the crucial physical quantities of heavy-ion collisions (HICs), and can affect obviously many observables at the final state, such as the multifragmentation and the collective flow.
 Usually, it cannot be measured directly in experiments but might be inferred from observables at the final state. Artificial intelligence has had great success in learning complex representations of data, which enables novel modeling and data processing approaches in physical sciences. In this article, we employ two of commonly used algorithms in the field of artificial intelligence, the Convolutional Neural Networks (CNN) and Light Gradient Boosting Machine (LightGBM), to improve the accuracy of determining impact parameter by analyzing the proton spectra in
transverse momentum and rapidity on the event-by-event basis. Au+Au collisions with the impact parameter of 0$\leq$$b$$\leq$10 fm at intermediate energies ($E_{\rm lab}$=$0.2$-$1.0$~GeV$/$nucleon) are simulated with the ultrarelativistic quantum molecular dynamics (UrQMD) model to generate the proton spectra data. It is found that the average difference between the true impact parameter and the estimated one can be smaller than 0.1 fm. The LightGBM algorithm shows an improved performance with respect to the CNN on the task in this work. By using the LightGBM's visualization algorithm, one can obtain the important feature map of the distribution of transverse momentum and rapidity, which may be helpful in inferring the impact parameter or centrality in heavy-ion experiments.
\end{abstract}



\maketitle


\section{Introduction}\label{section1}
Nuclear matter with high density and isospin asymmetry can be created during heavy-ion collisions (HICs). For example, in HICs at intermediate energies (several hundreds MeV/nucleons, i.e, beam energies available at facilities on the FRIB in the United States, the RIBF at RIKEN in Japan, the SPIRAL2 at GANIL in France, the CSR at HIRFL and the HIAF in China, the FAIR at GSI in Germany),  nuclear matter with about 2-3 times of the saturation density can be formed. Studying the property (e.g., the equation of state) of this dense state is one of hot topics in both nuclear physics and astrophysics \cite{BALi,Tsang,XuJun,Oertel:2016bki,Li:2018lpy,Li:2019xxz,Colonna:2020euy,Ma:2018wtw,Ono:2019jxm,LLV}. It is easy to conjecture that the state of the created dense nuclear matter depends on the colliding system (projectile and target), beam energy, and the impact parameter. The colliding system and beam energy can be set in heavy-ion experiments, however determining the impact parameter or the centrality is still a challenge for experimental techniques, as its typical value is only a few femtometers. Usually, the centrality of a collision is inferred from the observables at the final state. For example, the total charged multiplicity, the total transverse kinetic energy of light charged particles, the ratio of transverse-to-longitudinal kinetic energy, and the number of participating nucleons have been used to determine the centrality in experiments \cite{JGosset,YXZPRC,LFPLB,Andronic,lpcjpg,LLprc,hades}.

 Since 1990s, machine learning algorithms, such as support vector machine (SVM) and neural network (NN), have been used to determine the impact parameter \cite{SBASS,Haddad,JD,CDPRC}, the average difference between the estimated and the true impact parameter was found to be about 0.2 fm. Some statistical learning frameworks, such as Bayesian methods, also have been widely applied in nuclear physics \cite{PJC,PLB1,PLB2,MACPC,NIUPRC,XWJAPJ,XJARX,YePRL,PMPLB,JEPLB,LHLEPJA}. Recently, the field of artificial intelligence (AI) has received unprecedented attention, and prodigious progress has been made in the application of the AI techniques \cite{Lecun,CNN,RNN,GAN}. Particularly, the neural network framework that is based on the simulation of the brain has been widely used in many physical researches \cite{PLG,XLF,HL,ROMP,Nature1,sci1,NP,EPJC,JAN,DYLEPJC,PRD,PRDZHOUKAI,EOS,PRL124,XRL}. Thus, it is timely to deduce the impact parameter with the new developed machine learning and deep learning algorithms. In this paper, the Convolutional Neural Networks (CNN) and Light Gradient Boosting Machine (LightGBM) are used to determine the impact parameter from the two-dimensional transverse momentum and rapidity spectra of protons on the event-by-event basis.

The paper is organized as follows:
 In Sec.\uppercase\expandafter{\romannumeral2}, we will briefly introduce the UrQMD model and the generated two-dimensional transverse momentum and rapidity spectra.
 We continue with Sec.\uppercase\expandafter{\romannumeral3} in which CNN and LightGBM algorithms will be briefly described.
 In Sec.\uppercase\expandafter{\romannumeral4}, we discuss the results obtained by these two algorithms.
 We end with Sec.\uppercase\expandafter{\romannumeral5} dedicated to summary and outlook.
\section{UrQMD Model}\label{section2}
The Ultrarelativistic Quantum Molecular Dynamics (UrQMD) model is a many-body microscopic transport model which has been widely applied to simulate HICs in a very broad energy range \cite{SAB,BLE,qfli1,qfli2,FOP-wyj,FOP-zyx}.
In the UrQMD model, each hardon is represented by the Gaussian wave packet in phase space. The coordinate $r_{i}$ and momentum $p_{i}$ of particles \emph{i} are propagated according to Hamilton's equation of motion:
\begin{eqnarray}
\dot{\textbf{r}}_{i}=\frac{\partial  \langle H  \rangle}{\partial\textbf{p}_{i}},
\dot{\textbf{p}}_{i}=-\frac{\partial  \langle H \rangle}{\partial \textbf{r}_{i}}.
\end{eqnarray}

Here, {\it $\langle H \rangle$} is the total Hamiltonian function, it consists of the kinetic energy $T$ and the effective interaction potential energy $V$. For studying HICs at intermediate energies, the following density and momentum dependent potential has been widely employed in QMD-like models \cite{AIC,qfli3,CH,TLY,YU,feng},
\begin{equation}\label{eq2}
V=\alpha\left(\frac{\rho}{\rho_0}\right)+\beta\left(\frac{\rho}{\rho_0}\right)^{\gamma} + t_{md} \ln^2[1+a_{md}(\textbf{p}_{i}-\textbf{p}_{j})^2]\frac{\rho}{\rho_0}.
\end{equation}

Here $\alpha$=-390 MeV, $\beta$=320 MeV, $\gamma$=1.14, $t_{md}$=1.57 MeV, and $a_{md}$=500 $c^{2}$/GeV$^{2}$. With the above parameter set, a soft and momentum dependent (SM) equation of state (EoS) with the incompressibility $K_0$=200 MeV can be obtained. With $\alpha$=-130 MeV, $\beta$=59 MeV, and $\gamma$=2.09, a hard and momentum dependent (HM) equation of state with the incompressibility $K_0$=380 MeV can be yielded. In addition, the Coulomb potential for charged particles and the symmetry potential for nucleons are also included.
The pauli blocking and the in-medium nucleon-nucleon cross section are set in the same way as our previous studies, where a large amount experimental data from HICs at SIS energies has been reproduced with the UrQMD model \cite{Li:2018bus,Wang:2018hsw,wyj-plb2,Du:2018ruo,Liu:2018xvd}.

It is known that essentially the transverse momentum $p_{t}$=$\sqrt{p_x^2+p_y^2}$ and rapidity $y_{z}$=$\frac{1}{2}\ln[\frac{E+p_{z}}{E-p_{z}}]$ of charged particles can be measured in heavy-ion experiments. Throughout, the reduced rapidity $y_{0}$=$y_z$/$y_{pro}$ is used instead of $y_z$, in the same way as done in the experimental report \cite{fopi}. Here, $y_{pro}$ denotes the rapidity of the projectile in the c.o.m system. The range of $p_{t}$ is from 0 to 1 GeV/c and $y_{0}$ is from -2 to 2, in this case, more than 99\% protons can be included. To assess the accuracy of the reconstruction of impact parameter, the performance of different algorithms can be quantified by \cite{SBASS}:
  \begin{equation}\label{4}
  \Delta b = \frac{1}{N_{event}}\sum_{i=1}^{N_{events}}|b_{i}^{true}-b_{i}^{pred}|.
  \end{equation}
 Here, $b_i^{true}$ is the true impact parameter of each event, $b_i^{pred}$ is the predicted value from CNN and LightGBM algorithms.

\section{CNN and LightGBM algorithms}\label{section3}

The determination of impact parameter is set as a regression task in this work, the input is the two-dimensional transverse momentum and rapidity spectra of all protons simulated with the UrQMD model, and the output is the predicted impact parameter. To study the influence of the grid of the two-dimensional spectra, the transverse momentum and rapidity distributions can be divided into several bins, the smallest input grid is $3\times3$ and the largest one is $40\times40$ in this work. We note here that, detailed information of the spectra may lose with small grid, while with larger grid, event-by-event fluctuation of the spectra may conceal other effects.

\begin{figure*}[htbp]
\centering
\includegraphics[scale=0.5]{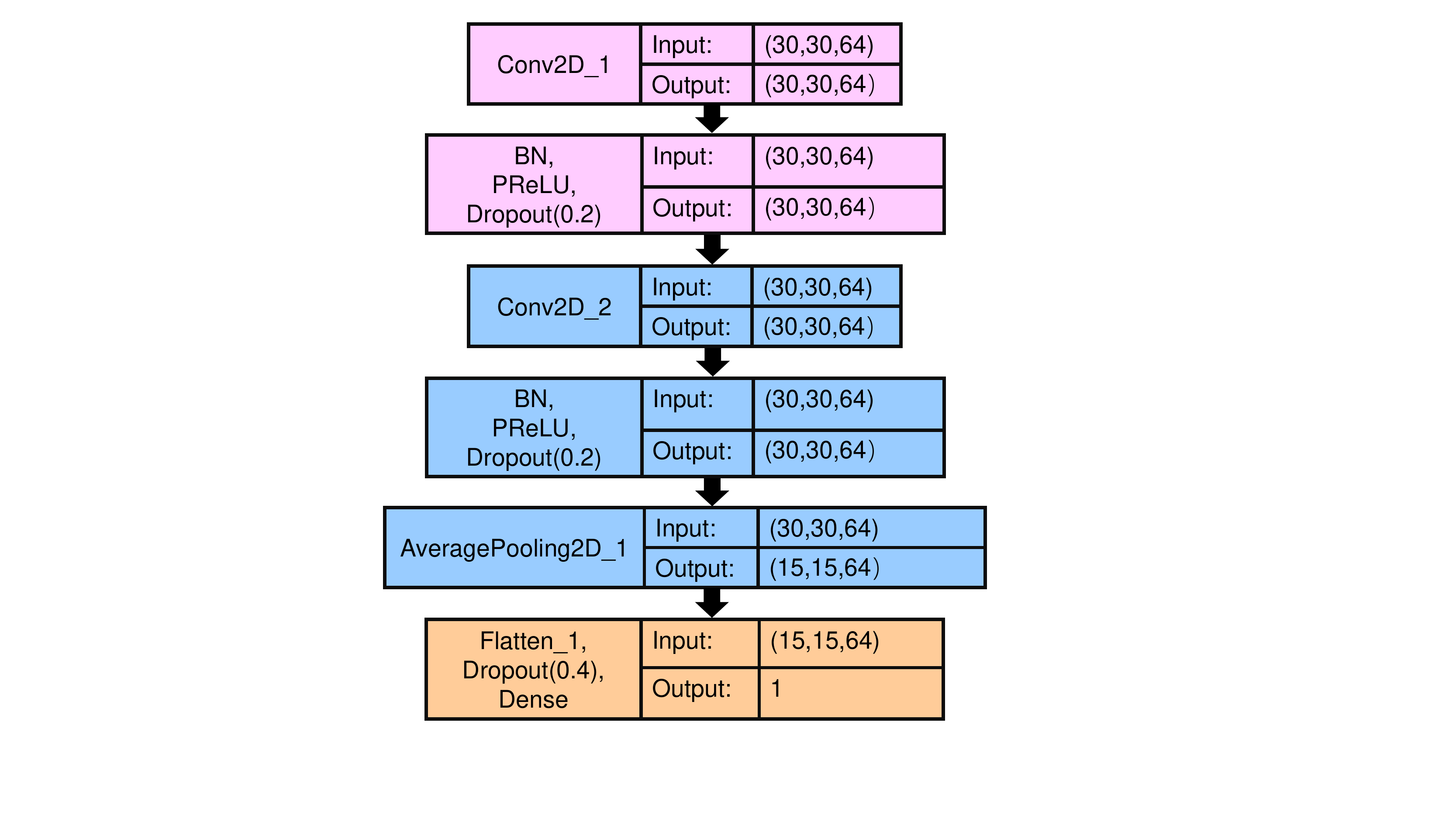}\\
\caption {\label{fig1} The architecture of convolution neural network used in this article. The architecture is designed to predict impact parameter by using $30\times30$ as the input grid.}
\end{figure*}

CNN refers to a powerful variant of neural networks, where the input into each of the hidden units is obtained using a filter applied to a small part of the input space. Because of the invariance to translation, CNNs are particularly suitable for image processing. Compared to the fully connected neural networks, each layer of CNN has much smaller number of parameters \cite{CNN1,CNN2}. It usually consists of the input layer, convolutional layer, pooling layer, fully connected layer and output layer. The input layer is used for retaining the structure and information of the input data for feature extraction. Then in the convolution layer, there are several important concepts, such as local receptive fields and shared weights that need to be trained. A local receptive field with kernel will generate a feature map by scanning and filtering the data. Between convolutional layer and pooling layer, usually batch normalization (BN), dropout and activation are added to prevent the gradient dispersion and overfitting \cite{BN,DP,AC}. The purpose of the pooling layer is to reduce the spatial dimensions. Before entering the final output layer, fully-connected layer is often used to transform the previously extracted local features into global features. After a series of analyses, the output layer returns the final predicted value. To minimize the error between the predicted value and the true value, a commonly used optimization method is the stochastic gradient descent (SGD). The CNN architecture used in this work is shown in Fig. \ref{fig1}. We use two convolution layers and one subsequent fully-connected layer. Each convolution layer is followed by the BN, parametric rectified linear unit (PReLU), dropout (with a rate of 0.2) and average pooling layer (of pool size $2\times2$, following the second convolutional layer only). There are 64 filters of size $5\times5$, in the two convolutional layers, scanning through the output from the previous layer and creating 64 feature maps of size $30\times30$ as input for the next layer. The resulting 64 feature maps of size $5\times5$ from the last average pooling layer are flattened. At the end of the fully-connected layer, a dropout with a rate of 0.4 is added. By setting the above parameters, a more stable and reliable CNN model can be obtained.

  LightGBM is a highly efficient algorithm based on gradient boosting decision tree (GBDT) for tackling the problem of high-dimensional features and large size data \cite{LGB}.
 The decision tree algorithm has widely used in the study of physical problems \cite{XIA,BPR,HJY}.
 LightGBM mainly includes two novel technologies: Gradient-based One-Side Sampling (GOSS) and Exclusive Feature Bundling (EFB).
 GOSS pays more attention to under-trained samples, and does not change the original data too much.
 As for EFB algorithm, it can avoid useless calculation of features and retain feature information as much as possible.
 The parameters of LightGBM, such as num\_leaves and max\_depth can be adjusted according to the actual situation.
 Here, the value of num\_leaves is 350, max\_depth is -1 and other related parameters are basically with their default values. Based on the preprocessed data, LightGBM uses histogram optimization algorithm to calculate the gain of each feature, and then sets a threshold for each feature. In each training, the result is continuously optimized through GOSS and EFB techniques. For more details, readers are referred to official documentation \cite{LightGBM}.
\section{Results and Discussions}\label{section4}

\begin{figure}[h]
\begin{centering}
\includegraphics[width=0.5 \textwidth]{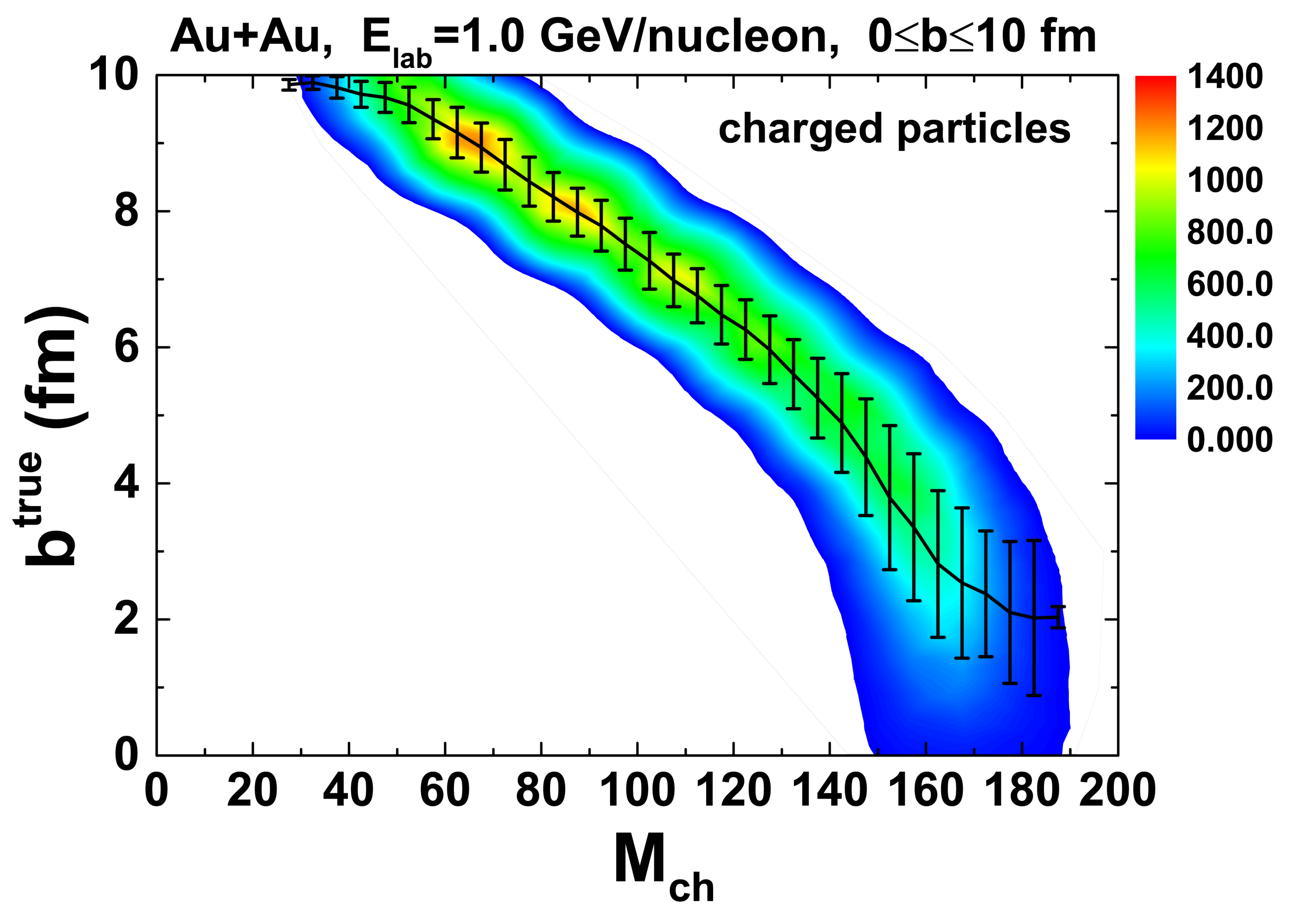}
\caption {\label{fig2} The correlation between the true impact parameter and the number of charged particles produced from Au+Au collisions at $E_{\rm lab}=1.0$~GeV$/$nucleon. The solid line and error bars represent the mean value of $b$ and its standard deviation in each $M_{ch}$ bin, respectively. Colors indicate the number of events within certain values of $M_{ch}$ and $b$. }
\end{centering}
\end{figure}

At present, different methods have been applied to deduce the collision centrality (or the impact parameter) in heavy-ion experiments \cite{Andronic,lpcjpg,LLprc,hades}. For example, the total charged multiplicity $M_{ch}$, total transverse kinetic energy of light charged particles $E_{\perp12}$, and the ratio of transverse-to-longitudinal kinetic energy $ERAT$ have been widely used to infer centrality in heavy-ion collision at intermediate energies \cite{lpcjpg}. It is not difficult to presume that the values of $M_{ch}$, $E_{\perp12}$ and $ERAT$ will be larger for collisions with smaller impact parameter. However, there is no strict one-to-one relationship between the impact parameter and these observables, the distribution of $M_{ch}$, $E_{\perp12}$ and $ERAT$ have been found to be smeared out, as illustrated in Fig. \ref{fig2} where the correlation of the total charged multiplicity $M_{ch}$ vs the impact parameter is displayed. The uncertainty of the inferred impact parameter solely from $M_{ch}$ is roughly equivalent to 3 fm due to the smearing effect.

\begin{figure}[h]
\begin{centering}
\includegraphics[width=0.5\textwidth]{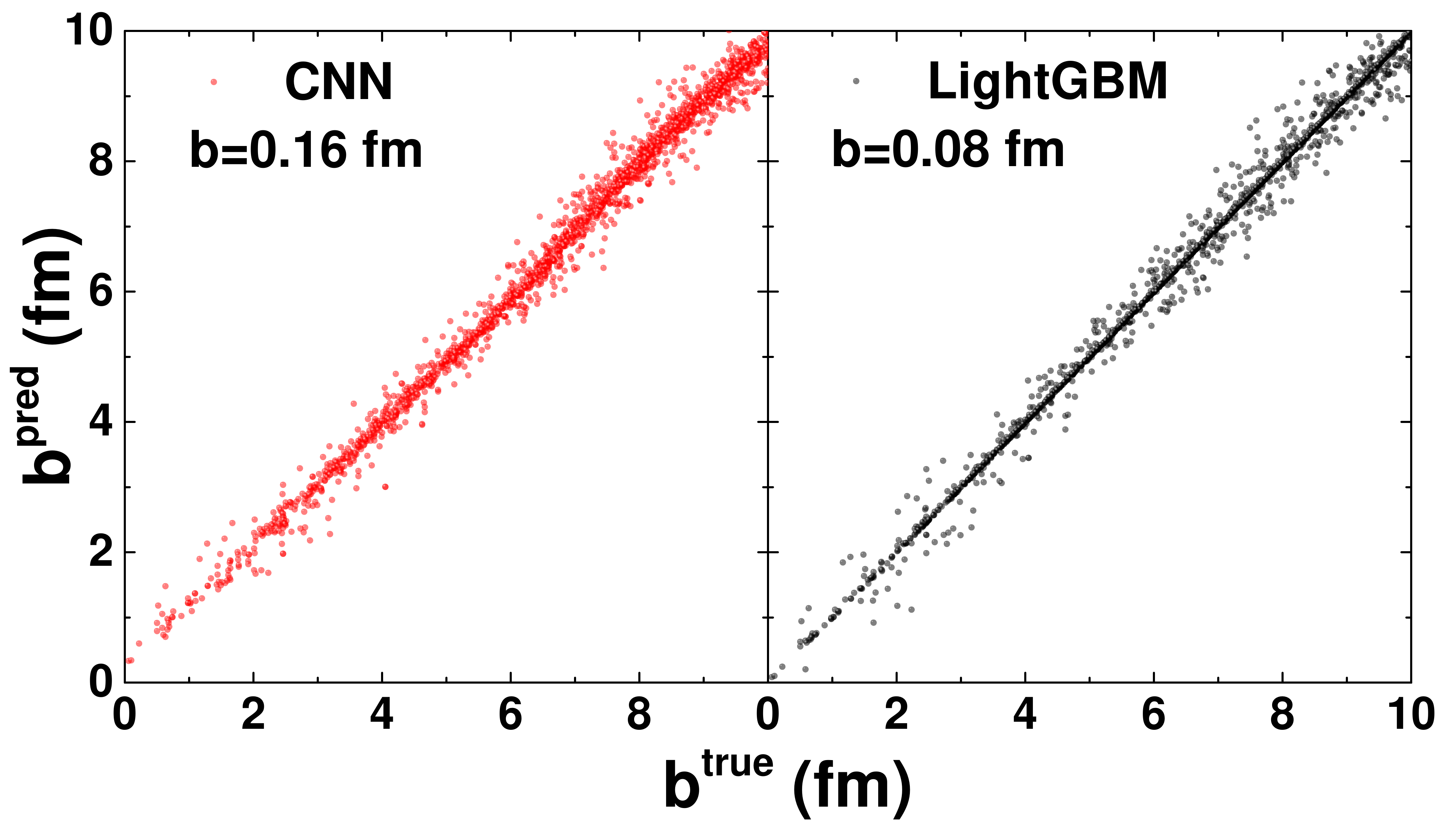}
\caption {\label{fig3} True impact parameter versus the predicted impact parameter for the 2000 testing data with CNN and LightGBM algorithms. The input grid $n^2$=$30\times30$ and the training data is taken from 50000 events. Each dot represents one Au+Au collision event at $E_{\rm lab}=1.0$~GeV$/$nucleon.}
\end{centering}
\centering
\end{figure}

To infer the impact parameter with AI, one needs to generate data. The data can be either from measurements in experiment or from simulations with theoretical models. In this work, we use the data generated with the UrQMD transport model. The data (i.e., the transverse momentum and rapidity spectra of protons on event-by-event basis) obtained from 50000 Au+Au collision events with known impact parameter 0$\leq$$b$$\leq$10 fm are set as the training data, and the data from another 2000 events are set as testing data. We have checked that with further increasing the testing data to 10000 events and dividing them into five sub-testing data samples, the standard deviation of the obtained $\Delta b$ is smaller than 0.02 fm, thus throughout 2000 Au+Au events are applied to validate the performance of AI algorithms on the determination of impact parameter, and the associated statistical uncertainty on the obtained $\Delta b$ is less than 0.02 fm.

\subsection{The training data size dependence}\label{section1}
The influence of the size of training data on $\Delta b$ is tested by setting the training data from 2000, 10000, 20000, 40000, and 50000 events. The results are displayed in Table. \ref{tab1}. The difference between the predicted and true impact parameter decreases roughly from $\Delta b$=0.3 fm to $\Delta b$=0.1 fm when we increase the size of the training data from 2000 to 50000 events. It is understandable that more detailed information can be found out with much more training data. We further increase the training data to 100000 events, and find that $\Delta b$ obtained with LightGBM decreases to 0.04 fm while it obtained with CNN nearly saturates at a constant value of 0.16 fm. Taking 50000 events as the training data, the obtained $\Delta b$ is already very small, thus throughout this work, the training data are generated from 50000 events. The good performance of CNN and LightGBM on the determination of impact parameter also can be easily observed in Fig. \ref{fig3} where the true impact
parameter is plotted versus the predicted impact parameter. One may also observed from Fig. \ref{fig3} that $\Delta b$ is somehow impact parameter dependent, i.e., $\Delta b$ is larger for smaller impact parameter. For example, the $\Delta b$=0.12 fm in 0$\le$$b$$\le$3 fm while $\Delta b$=0.06 fm in 3$\le$$b$$\le$6 fm with the LightGBM algorithm. It can be understood that, more random nucleon-nucleon collisions occur with a smaller impact parameter so that larger fluctuations take place for the proton spectra, which results in a nontrivial correlation between $b$ and the spectra.

It is known that there are always some cuts in experiment, as an example, we checked the influence of $p_t$ cut on the obtained $\Delta b$. It is found that, the obtained $\Delta b$ are about 0.22 and 0.29 fm when taken the data with $p_t$ below 0.2 and 0.4 GeV$/$c are cut off, respectively. While $\Delta b$ is about 0.08 fm for data without any $p_t$ cut. It is intuitively understandable that the obtained $\Delta b$ becomes larger when some information on the data are cut off.

\begin{table}[h]
\centering
\caption{\label{tab1} The dependence of $\Delta b$ on the number of training data at $E_{\rm lab}=1.0$~GeV$/$nucleon. The results are obtained with the input grid $n^2$=$30\times30$. }
\setlength{\tabcolsep}{1.3mm}
\begin{tabular}{cccccccc}
\hline
Size of training data         &2000    & 10000     & 20000   &40000    & 50000     \\ \hline
CNN                           &0.26    & 0.26     & 0.24   &0.16    & 0.16    \\ \hline
LightGBM                      &0.29    & 0.25     & 0.17   &0.08    & 0.08     \\ \hline
\end{tabular}
\end{table}
\
\subsection{Input grid dependence}\label{section2}

\begin{figure}[h]
\begin{centering}
\includegraphics[width=0.46\textwidth]{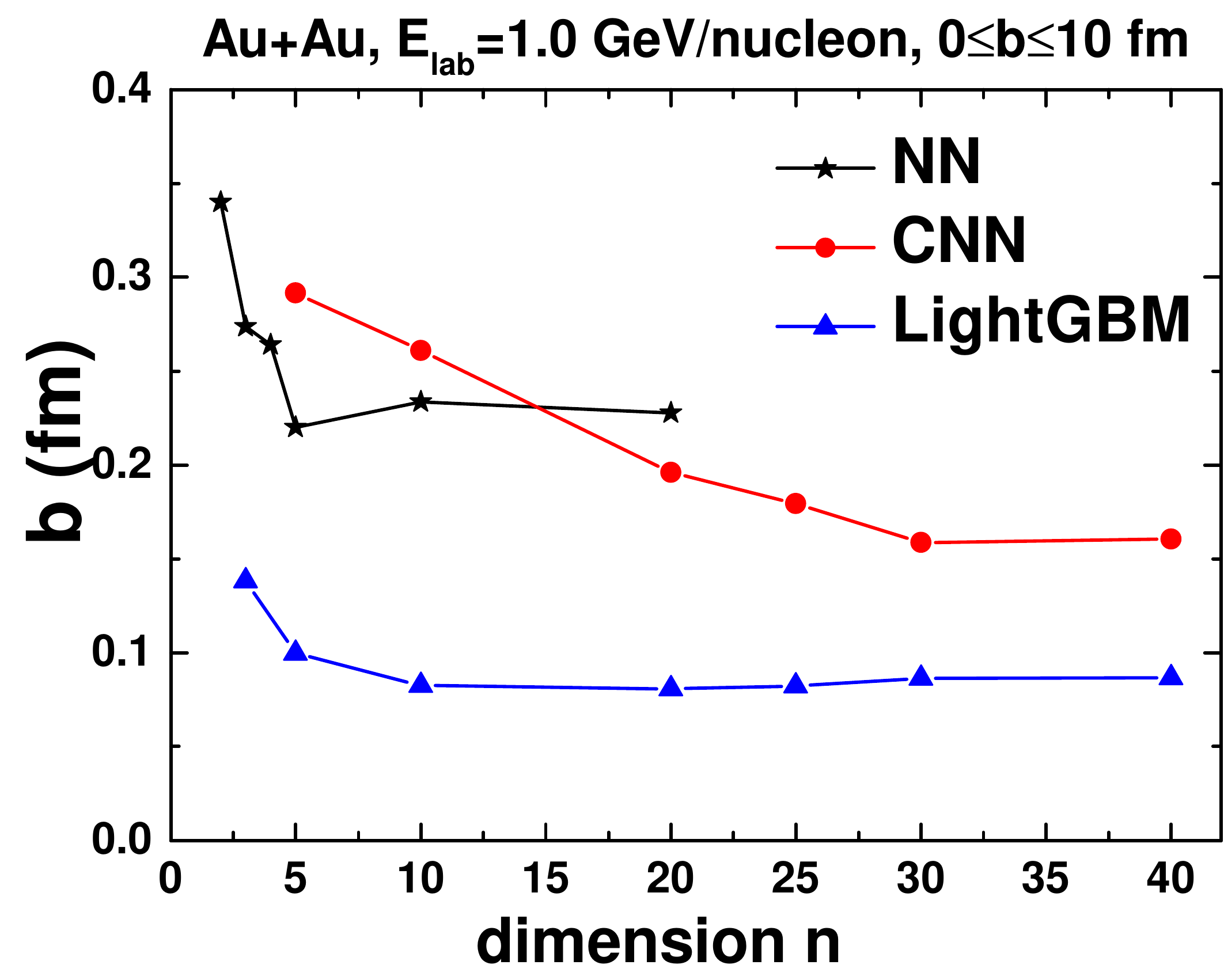}
\caption {\label{fig4}The $\Delta b$ dependence of the input grid dimension n. The results are obtained with the training data obtained from 50000 Au+Au events at E$_{\rm lab}$=1.0 GeV$/$nucleon. The result of NN are taken from Ref. \cite{SBASS}.}
\end{centering}
\end{figure}

We investigate how the obtained $\Delta b$ depends on the input grid dimension. We mentioned in the previous section that the different grid dimension may affect the obtained results, as displayed in Fig. \ref{fig4} where $\Delta b$ is plotted as a function of the input grid dimension $n\times n$. The obtained $\Delta b$ decreases with increasing the input grid dimension. Because more information can be provided with a larger grid dimension. The obtained $\Delta b$ almost maintains a constant value after $n\times n$ larger than $10\times10$ for LightGBM and larger than $30\times30$ for CNN. Very similar result also can be found in Ref. \cite{SBASS} where the NN was applied. $\Delta b$ obtained with LightGBM is always smaller than that obtained with CNN and NN. The CNN shows a better performance than NN only when the input grid $n\times n$ is larger than $20\times20$.

\subsection{Beam energy dependence}\label{section3}

The performance on the determination of impact parameter at beam energies of 0.2, 0.4, 0.6 and 0.8 GeV/nucleon is studied as well, and the results are listed in Table. \ref{tab3}. In general, the obtained $\Delta b$ increases with decreasing beam energy. This is partly due to the fact that the mean field potential effect is much more important at lower beam energies, it is hard to distinguish the participants (nucleons which suffered collisions) and spectators (nucleons which do not suffer any collision) as almost all nucleons can interact with each other by mean field potential. At higher beam energies, spectators will quickly fly away from the interacting region and less affect by participants, the information of the initial stage (e.g., the number of participant and spectator determined by the impact parameter) can be easier deduced from the observation at the final stage.
\begin{table}[h]
\centering
\caption{\label{tab3} The dependence of $\Delta b$ on the beam energy. The results are obtained with the training data obtained from 50000 events and the input grid dimension $n^2$=$30\times30$. }
\setlength{\tabcolsep}{1.3mm}
\begin{tabular}{cccccc}
\hline
$E_{\rm lab}$ (GeV$/$nucleon)         &0.2  & 0.4   & 0.6    &0.8     & 1.0   \\ \hline
CNN                             &0.43 & 0.31  & 0.15   & 0.15   & 0.16   \\ \hline
LightGBM                        &0.42 & 0.32  & 0.08  & 0.09  & 0.08  \\ \hline
\end{tabular}
\end{table}
\subsection{Model parameters dependence}\label{section4}
\begin{table}[h]
\caption{\label{tab4} The dependence of $\Delta b$ on various model parameters. The training data is generated with SM and $E_{\rm lab}$=1.0~GeV$/$nucleon. While the testing data are obtained with HM and at $E_{\rm lab}$=0.4, 0.6, 0.8, and 1.0~GeV$/$nucleon.}
\setlength{\tabcolsep}{1.3mm}
\begin{tabular}{|c|cc|cc|}
  \hline
  EoS & \multicolumn{2}{c|}{HM} & \multicolumn{2}{c|}{SM} \\
  \hline
   $E_{\rm lab}$~GeV/nucleon    & CNN & LightGBM & CNN & LightGBM \\
  \hline
  0.4 & 0.88 & 0.85 & 0.88 & 0.86 \\
  0.6 & 0.48 & 0.49 & 0.49 & 0.52 \\
  0.8 & 0.30 & 0.30 & 0.30 & 0.32 \\
  1.0 & 0.28 & 0.28 & 0.16 & 0.08 \\
  \hline
\end{tabular}
\end{table}
\begin{figure*}
\includegraphics[width=0.8\textwidth]{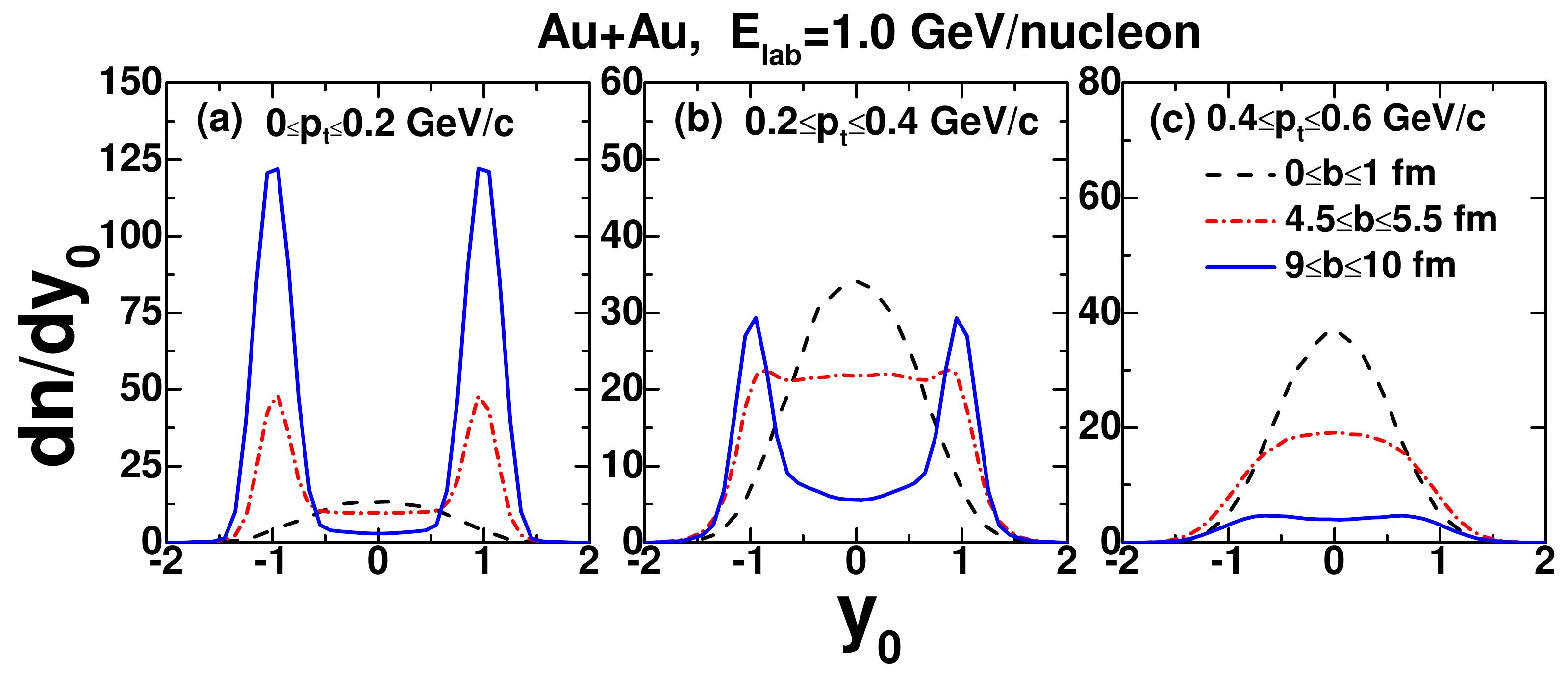}
\caption {\label{fig6} The rapidity distribution of protons from Au+Au at $E_{\rm lab}=1.0$~GeV$/$nucleon. From left to right, the transverse momentum cuts 0$\le$$p_t$$\le$0.2 GeV$/$c (a), 0.2$\le$$p_t$$\le$0.4 GeV$/$c (b), and 0.4$\le$$p_t$$\le$0.6 GeV$/$c (c) are applied.}
\end{figure*}

CNN and LightGBM models are trained to learn more information about the impact parameter from the training data, it is expected that the trained model is valid even when the testing data is generated from experiments or from other transport models. To verify the generalizability of the trained CNN and LightGBM models, the proton spectra generated with SM at $E_{\rm lab}=1.0$~GeV$/$nucleon is solely set as the training data, while the data generated with HM and at $E_{\rm lab}$=0.4, 0.6, 0.8, and 1.0~GeV$/$nucleon are set as the testing data. The corresponding $\Delta b$ are listed in Table \ref{tab4}. As we known, the proton spectra can be visibly affected by the nuclear equation of state and beam energy, while the obtained $\Delta b$ at $E_{\rm lab}$=0.4 and 1.0~GeV$/$nucleon are about 0.9 fm and 0.3 fm, respectively. Although this results are larger than that listed in Table \ref{tab4}, both models trained with CNN and LightGBM are found to be able to capture highly-correlated features for determining the impact parameter from proton spectra. The precision on the determination of impact parameter will decrease when model parameter dependence is considered. Nevertheless, this issue can be partly reduced by using more data generated with models under more assumptions for model parameters. It will be addressed in a future study.

\subsection{Interpretation of the LightGBM algorithm}\label{section5}
\begin{figure}[h]
\begin{centering}
\includegraphics[width=0.5\textwidth]{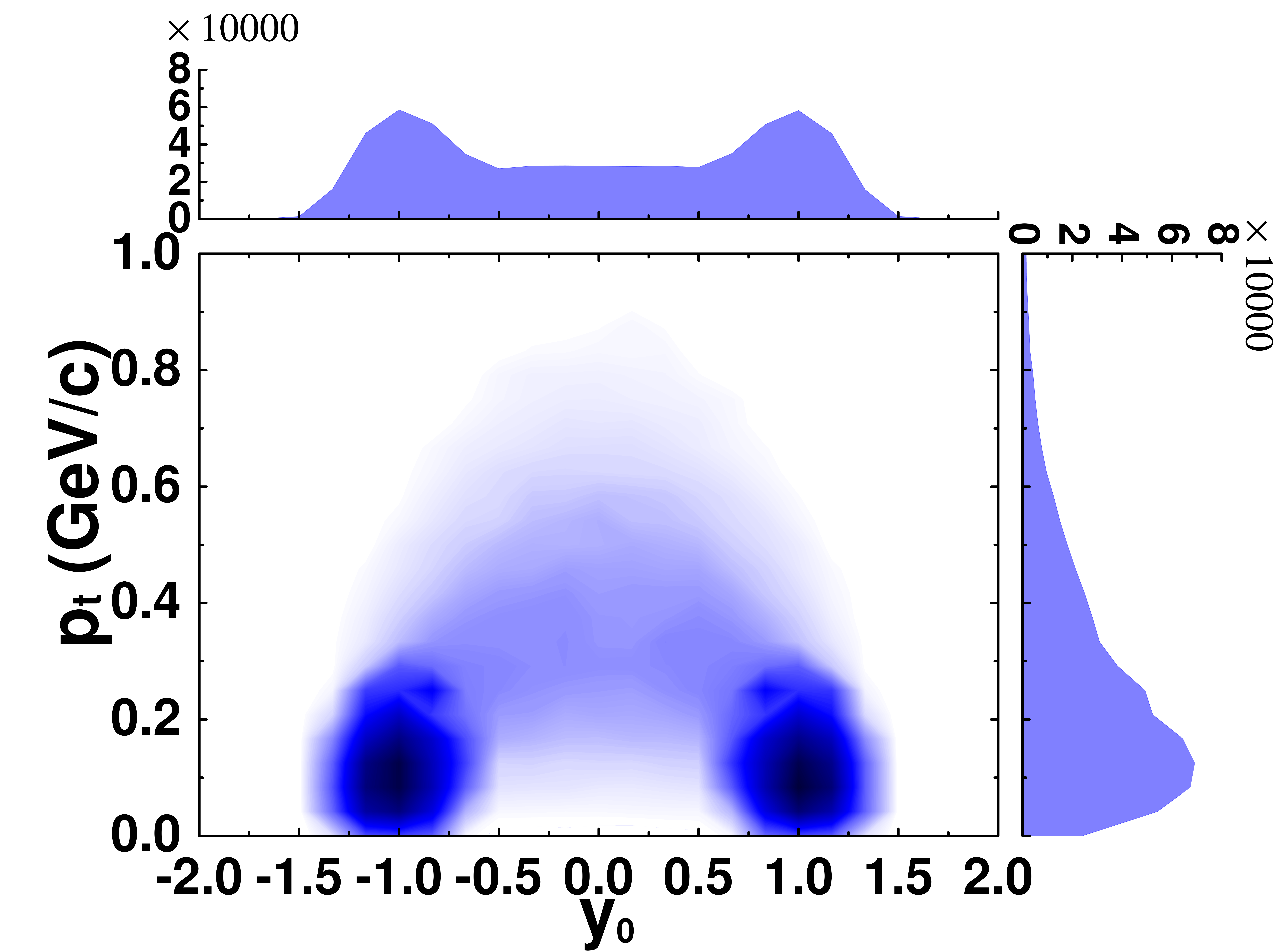}
\caption {\label{fig5}Importance map of the $p_{t}$-$y_{0}$ distribution obtained with the Feature\_importance technology of LightGBM algorithm. The depth of color represents the relative importance of each bin for determination of the impact parameter. }
\end{centering}
\end{figure}
We have seen the powerful ability of AI on the determination of impact parameter. One interesting question is that can we obtained some physical insights on the $p_{t}$-$y_{0}$ distribution
with artificial intelligence, and whether the obtained information can be understood with physical models. For this purpose, the Feature\_importance technology of LightGBM algorithm is applied to show which region of the $p_{t}$-$y_{0}$ distribution is more sensitive to the impact parameter. Fig. \ref{fig5} shows the importance map of the $p_{t}$-$y_{0}$ distribution. As one can see that, around $y_0$=$\pm$1.0 and $p_t$ $\le$ 0.2 GeV$/$c is the most important region to determine the impact parameter. Around the mid-rapidity $y_0$=0, it can be found that 0.2$\le$$p_t$$\le$ 0.6 GeV$/$c is also more important than other $p_t$ range. Shown in Fig. \ref{fig6} are the rapidity distribution of protons with different impact parameter in the windows 0$\le$$p_t$$\le$0.2 GeV$/$c (a), 0.2$\le$$p_t$$\le$0.4 GeV$/$c (b), and 0.4$\le$$p_t$$\le$0.6 GeV$/$c (c). For very peripheral collision (e.g., 9$\le$$b$$\le$10 fm), many protons will keep their initial longitudinal rapidity and transverse momentum, as a result two peaks can be observed around $y_0$=$\pm$1.0 for lower $p_t$ window. With a larger transverse momentum $p_t$ cut, the number of protons at mid-rapidity is larger for central collision than for peripheral collision. Therefore, these regions which are also found out with the Feature\_importance technology of LightGBM algorithm are more important in the determination of impact parameter.

\section{Summary and outlook}\label{section5}

In summary, two popular AI algorithms, CNN and LightGBM, are applied to determine the impact parameter from the final two-dimensional transverse momentum and rapidity spectra of protons on the event-by-event basis. By taking the spectra of 50000 Au+Au collision events as the training data, the average difference between the true impact parameter and the predicted one are found to be $\Delta b$=0.16 fm with CNN and $\Delta b$=0.08 fm with LightGBM algorithm for Au+Au collision at E$_{\rm lab}$=1.0~GeV$/$nucleon. The influence of the input grid and the size of training data on the performance are investigated, and it is found that $\Delta b$ roughly decreases with increasing input grid, $10\times10$ and $30\times30$ input grid are enough for LightGBM and CNN algorithms, respectively. In addition, the results at beam energies of 0.2, 0.4, 0.6, and 0.8 GeV$/$nucleon are investigated, and found that the obtained $\Delta b$ increases with decreasing beam energy. By varying model parameters, the generalizability of the trained models are tested as well. Furthermore, the importance map on the two-dimensional transverse momentum and rapidity spectra of protons can be obtained with LightGBM algorithm, and the most relevant regions extracted with LightGBM can be understood from the rapidity distribution of protons.

In this work, with the data generated from the UrQMD model, the possible application of artificial intelligence algorithms in studying heavy-ion collisions at intermediate energies has been demonstrated. The precision on the determination of impact parameter will decrease when model parameter dependence is considered, this will limit the usefulness of the AI methods in the study of heavy-ion collisions. To pursue a more promising avenue of research, on the data side, one may use data from different models and/or from experiments, then possible model-dependent contributions can be disentangled and more reliable information is expected to be deduced. On the AI algorithm side, it would be valuable and interesting to apply other methods, such as the Densely Connected Convolutional Networks (DenseNet) \cite{DenseNet}, Bayesian Neural Network (BNN) \cite{BNN}, eXtreme Gradient Boosting (XGBoost) \cite{XGB}, and some unsupervised learning method such as Sparse Autoencoder \cite{SP}, Restricted Boltzmann Machine (RBM) \cite{RBM}, Clustering algorithm \cite{CA}, to gain further insight from theoretical and experimental data as well as to obtain uncertainty estimates of physical variables.


\begin{acknowledgments}
Fruitful discussions with Jan Steinheimer, Yilun Du, Kai Zhou, Nan Su, Long-gang Pang, and Horst St\"{o}cker are greatly appreciated.
The authors acknowledge
support by the computing server C3S2 in Huzhou University. The work is supported in part by the National Natural
Science Foundation of China (Nos. 11875125 and 11947410), and the Zhejiang Provincial
Natural Science Foundation of China under Grants No. LY18A050002 and No. LY19A050001, and the ``Ten Thousand Talent Program" of Zhejiang province.

\end{acknowledgments}

\end{document}